\documentclass[
reprint,
superscriptaddress,
amsmath,
amssymb,
aps,
prx,
twocolumn]{revtex4-2}
\usepackage{graphicx}
\usepackage{dcolumn}
\usepackage{bbm}
\usepackage{physics}
\usepackage{csquotes}
\usepackage[title]{appendix}
\usepackage[normalem]{ulem}
\usepackage{comment}
\usepackage{bm}
\usepackage{subfigure}
\usepackage[colorlinks,citecolor=blue]{hyperref}
\usepackage{cleveref}
\usepackage{comment}
\usepackage{todonotes}

\usepackage{booktabs}
\usepackage{mathtools}
\usepackage{orcidlink}

\renewcommand{\Re}{\operatorname{Re}}
\renewcommand{\Im}{\operatorname{Im}}

\begin{document}


\title{Learning Pole Structures of Hadronic States using Predictive Uncertainty Estimation}

\newcommand{\aqa}{$\langle aQa ^L\rangle $ Applied Quantum Algorithms, Universiteit Leiden}
\newcommand{\lorentz}{Instituut-Lorentz, Universiteit Leiden, Niels Bohrweg 2, 2333 CA Leiden, Netherlands}
\newcommand{\liacs}{LIACS, Universiteit Leiden, Niels Bohrweg 1, 2333 CA Leiden, Netherlands}
\newcommand{\ulm}{Institute for Complex Quantum Systems, Ulm University, 89069 Ulm, Germany}
\newcommand{\iqst}{Center for Integrated Quantum Science and Technology (IQST), Ulm-Stuttgart, Germany}
\newcommand{\upp}{National Institute of Physics, University of the Philippines Diliman, Quezon City 1101, Philippines}


\author{Felix Frohnert \orcidlink{0000-0003-3717-6352}}
\thanks{These two authors contributed equally.}
\affiliation{\liacs}
\affiliation{\aqa}
\author{Denny Lane B. Sombillo \orcidlink{0000-0001-9357-7236}}
\thanks{These two authors contributed equally.}
\affiliation{\upp}
\email{dbsombillo@up.edu.ph}
\author{Evert van Nieuwenburg \orcidlink{0000-0003-0323-0031}}
\affiliation{\liacs}
\affiliation{\aqa}
\author{Patrick Emonts \orcidlink{0000-0002-7274-4071}}
\email{patrick.emonts@uni-ulm.de}
\affiliation{\ulm}
\affiliation{\iqst}
\affiliation{\lorentz}
\affiliation{\aqa}

\date{\today}

\begin{abstract}
Matching theoretical predictions to experimental data remains a central challenge in hadron spectroscopy. 
In particular, the identification of new hadronic states is difficult, as exotic signals near threshold can arise from a variety of physical mechanisms. 
A key diagnostic in this context is the pole structure of the scattering amplitude, but different configurations can produce similar signatures. 
The mapping between pole configurations and line shapes is especially ambiguous near the mass threshold, where analytic control is limited.
In this work, we introduce an uncertainty-aware machine learning approach for classifying pole structures in $S$-matrix elements. 
Our method is based on an ensemble of classifier chains that provide both epistemic and aleatoric uncertainty estimates.
We apply a rejection criterion based on predictive uncertainty, achieving a validation accuracy of nearly $95\%$ while discarding only a small fraction of high-uncertainty predictions.
Trained on synthetic data with known pole structures, the model generalizes to previously unseen experimental data, including enhancements associated with the $P_{c\bar{c}}(4312)^+$ state observed by LHCb. 
In this, we infer a four-pole structure, representing the presence of a genuine compact pentaquark in the presence of a higher channel virtual state pole with non-vanishing width. 
While evaluated on this particular state, our framework is broadly applicable to other candidate hadronic states and offers a scalable tool for pole structure inference in scattering amplitudes.
\end{abstract}
\maketitle

\section{Introduction}
\label{sec:introduction}

\begin{figure*}[t]
    \centering
    \includegraphics[width=1\linewidth]{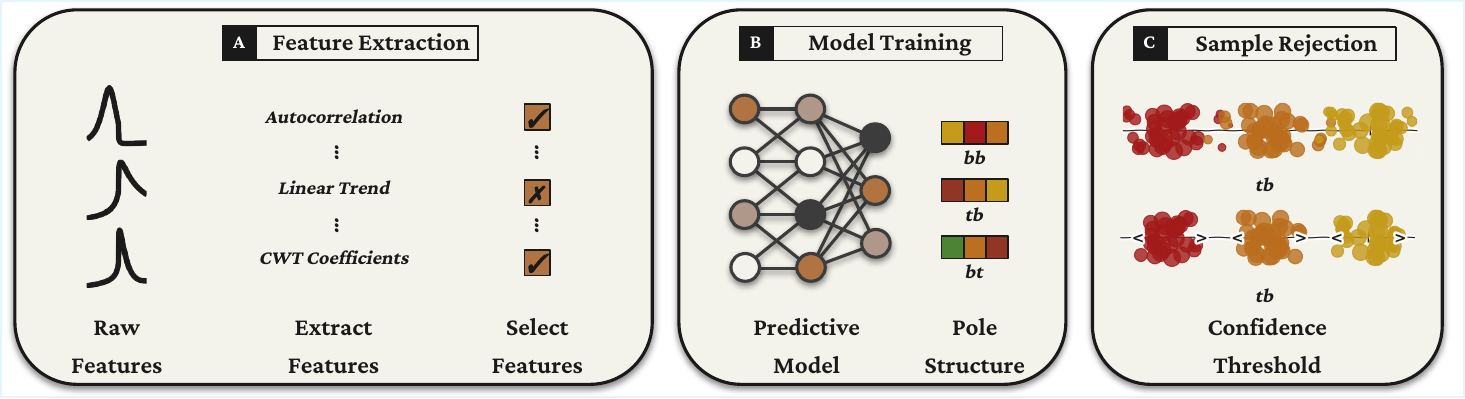}
    \caption{\textbf{Overview: } 
    Workflow for learning pole structures using predictive uncertainty estimation.
(a) Time-series features are extracted from raw line shapes, and a subset of relevant features is selected for model training.
(b) A predictive model is trained to infer pole structures based on the selected features.
(c) Model uncertainty is leveraged to filter predictions: all predictions are initially visualized (top), with sphere size indicating model confidence.
Predictions below a chosen confidence threshold are discarded, reducing outliers and increasing reliability.
    }
    \label{fig:overview}
\end{figure*}

Recent observations of near-threshold phenomena by different experimental collaborations present an opportunity to deepen our understanding of the non-perturbative nature of strong interaction dynamics~\cite{Olsen:2017bmm,LHCb-FIGURE-2021-001-report,Liu:2023hhl,Mai:2022eur}. 
Candidate hadronic states located below certain two-hadron thresholds are typically associated with the formation of hadronic molecules~\cite{Guo:2017jvc}. 
However, such a straightforward interpretation remains contested. 
In particular, some studies suggest that the observed exotic signals may originate from multiquark states beyond the conventional quark model~\cite{Chen:2022asf,Ali:2019roi}. 
Other competing interpretations are entirely kinematical, such as threshold cusps or triangle singularities~\cite{Guo:2019twa}, and do not require the formation of unstable quantum states. 
Once these kinematical effects are unambiguously ruled out, the next step is to determine the nature of the unstable hadronic state.

Such candidate hadrons manifest as pole singularities in the scattering amplitude, and their distribution across different Riemann sheets encodes essential information about their internal structure~\cite{Morgan:1990ct,Morgan:1992ge,Badalian:1981xj}. 
The identification of these pole structures can, in principle, be performed in a model-independent manner. 
However, different pole configurations can produce nearly indistinguishable line shapes in the invariant mass distributions of the final-state particles. 
For instance, an enhancement resulting from a hadronic molecule may closely resemble that of a genuine resonance whose profile is modified by coupled-channel effects. 
As a result, conventional fitting procedures often yield ambiguous conclusions, particularly when complementary data from other processes is unavailable. 
To address this challenge, alternative approaches beyond conventional amplitude fits are required to distinguish such subtle scenarios.

Machine learning (ML) has recently emerged as a promising tool in hadron spectroscopy, offering new strategies to resolve these ambiguities. 
Neural networks have demonstrated the ability to classify pole types and spin-parity assignments directly from synthetic or experimental line shapes~\cite{sombilloModelIndependentAnalysis2021,zhangRevealingNatureHidden2023,coAnalysisHiddencharmPentaquarks2024}. 
However, many existing approaches share one central limitation: the absence of rigorous uncertainty quantification.

In this work, we develop a machine learning framework that directly addresses this shortcoming. 
Synthetic line shapes are generated using a fully analytic, coupled-channel $S$-matrix model~\cite{sombilloModelIndependentAnalysis2021}. 
Pole configurations are constructed independently using a uniformized variable formalism, enabling precise control over the number and placement of singularities across Riemann sheets. 
This model-agnostic setup avoids assumptions about pole trajectories and enables the construction of a balanced, diverse training dataset.

We design a machine learning pipeline that optimizes all critical stages of the learning process: from data representation to confidence-based sample rejection, as visualized in Fig.~\ref{fig:overview}. 
An ensemble of Gradient Boosting~\cite{natekinGradientBoostingMachines2013,dorogushCatBoostGradientBoosting2018} classifier chains~\cite{readClassifierChainsMultilabel2011,readClassifierChainsReview2021,zhangReviewMultiLabelLearning2014} is trained to classify pole structures from line shapes while also quantifying uncertainty via posterior sampling.

We begin by constructing a set of informative input features and formulate an efficient representation of the learning task. 
We then perform extensive model selection to identify the architecture that yields the best learning performance. 
Finally, we show how predictive uncertainty can be used for confidence-based sample selection, enabling the model to reject unreliable predictions and improve overall robustness.

We apply this framework to the $P_{c\bar{c}}(4312)^+$ enhancement and find that the line shape is most consistent with the pole configuration $[bt]=[1]$, $[bb]=[2]$, and $[tb]=[1]$; a novel result in machine-learned line shape analysis.
Here, $[bt]$, $[bb]$, and $[tb]$ are the names of different Riemann sheets which are detailed below.
This structure supports the interpretation of a compact hidden-charm pentaquark, consistent with the GlueX observations reported in~\cite{GlueX:2023pev,Strakovsky:2023kqu}. 
Our analysis shows that this configuration is favored with high model-based confidence, while alternative interpretations receive only marginal support. 
By combining rigorous $S$-matrix modeling with uncertainty-aware machine learning, our approach offers a powerful and generalizable tool for classifying resonance structures in hadron spectroscopy.

The remainder of this manuscript is organized as follows: 
Section~\ref{sec:theory} outlines the theoretical framework underlying the scattering analysis and $S$-matrix parameterization. 
Section~\ref{sec:algorithm} introduces the machine learning pipeline, including the formulation of the learning task, model selection, and prediction of uncertainties. 
Section~\ref{sec:results} presents our analysis of the $P_{c\bar{c}}(4312)^+$ line shape and the inferred pole structures. 
Finally, Section~\ref{sec:conclusion} summarizes the key findings and discusses future directions.

\section{Theoretical Background}
\label{sec:theory}
Our analysis of near-threshold line shapes is based on coupled-channel scattering theory. 
We begin by reviewing the analytic structure of the scattering amplitude in multi-channel systems, focusing on the role of Riemann sheets and the interpretation of resonance poles.
By introducing a uniformization procedure, we provide a compact representation of the coupled-channel $S$-matrix in terms of a single complex variable, making the classification of pole configurations more tractable.
We parameterize the $S$-matrix using Jost-like functions in a flexible and physically consistent manner, which allows for the independent placement of poles across different sheets while preserving unitarity and analyticity.
To generate training data, we apply this framework to the case of the $P_{c\bar{c}}(4312)^+$, outlining the generation of synthetic line shapes from model amplitudes and the labeling of pole configurations that form the basis for the machine learning task.

\begin{figure*}[t!]
    \centering
    \includegraphics[width=\linewidth]{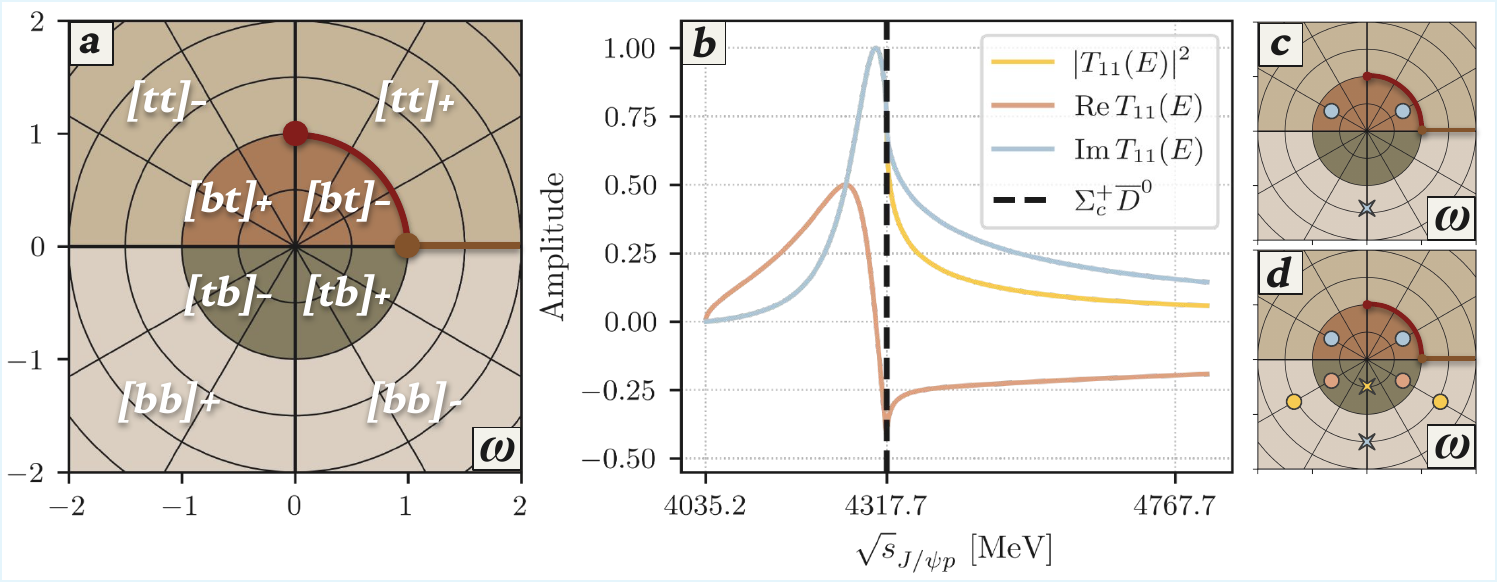}
    \caption[]{\textbf{Coupled-channel scattering overview:}
(a) Regions in the $\omega$-plane mapped to distinct energy Riemann sheets. 
Subscripts $+$ and $-$ indicate the upper and lower halves of each sheet, respectively. 
The points $\omega = i$ and $\omega = 1$ correspond to the thresholds $\epsilon_1$ and $\epsilon_2$. 
The solid arc (red) of the unit circle maps to the scattering region between these thresholds, while the solid segment along the real axis beyond $\omega = 1$ maps to the region above the second threshold.
(b) The line shape, showing a peak below the second threshold, can be reproduced by either (c) a single pole located in the $[bt]$ sheet or (d) a configuration involving three poles.}\label{fig:ambiguity}
\end{figure*}

\subsection{Coupled-channel scattering}
The analytic structure of scattering amplitudes is characterized by two fundamental types of singularities: 
poles and branch points. 
Bound states correspond to the discrete spectrum of the system’s Hamiltonian and manifest as simple poles of the scattering amplitude located below the lowest threshold. Branch points, on the other hand, are associated with scattering thresholds and arise from the multi-valued nature of the amplitude along the continuous spectrum of scattering energy. These singularities directly manifest in the Green's operator associated with the two-particle Hamiltonian. Specifically, for a spherically symmetric Hamiltonian, where $(\ell,m)$ are good quantum numbers, the Green's operator takes the form
\begin{equation}
    \begin{split}
    \hat{G}(z)&=\left(z-\hat{H}\right)^{-1} \\
    &=\sum_n\dfrac{|n\rangle\langle n|}{z-E_n}
    +\int_{E_{th}}^{\infty}dE\sum_{\ell,m}
    \dfrac{|E,\ell,m\rangle\langle E,\ell,m|}{z-E}
    \end{split}
    \label{eq:green}
\end{equation}
where $\{|n\rangle\}_n$ forms the discrete spectrum and $\{|E,\ell,m\rangle\}_{E,\ell,m}$ for the continuous spectrum. The first term contains simple poles in the complex $z$ variable while the second term will result into multi-valuedness when $z$ approched the real energy axis above the threshold, $E>E_{th}$. These singularities of the Green's operator are inherited by the scattering amplitude via the Lippmann-Schwinger equation.

The nature of branch point singularities is clarified by invoking the two-body unitarity condition of the scattering process, which implies that the reciprocal of the scattering amplitude is proportional to the two-body phase space. 
Since the phase space depends on the momentum, which scales as the square root of the scattering energy, the singularity appears as a square-root branch point.
More precisely, in a two-particle system, the invariant mass $\sqrt{s}$ relates to the center-of-mass momentum of channel $i$ as
\begin{equation}
    p_i = \frac{\sqrt{s - \epsilon_i^2}\sqrt{s - \epsilon_i(\epsilon_i - 4\mu_i)}}{2\sqrt{s}},
\end{equation}
where $\epsilon_i$ is the threshold energy of channel $i$ and $\mu_i$ is the reduced mass. 
The scattering region is defined by $\sqrt{s} \geq \epsilon_i$, placing the square-root branch point at the threshold $\sqrt{s} = \epsilon_i$. 
Near this threshold, it is convenient to define $q_i \propto \sqrt{s - \epsilon_i^2}$, which captures the correct analytic behavior of the amplitude in the vicinity of the singularity.

In the single-channel case, the complex $q_i$ plane is divided into upper and lower half-planes. 
The upper half-plane corresponds to the physical Riemann sheet, which contains the scattering region. 
Bound-state poles reside on the upper imaginary axis of the momentum plane, as their associated wave functions are normalizable. 
Due to causality, the physical sheet must be free of singularities, aside from bound-state poles below the lowest threshold~\cite{vankampen$S$MatrixCausalityCondition1953,vankampen$S$MatrixCausality1953}.
In contrast, the lower half-plane defines the unphysical Riemann sheet, which is not constrained by analyticity. 
Virtual poles appear on the lower half of the imaginary axis and are not associated with physical states. 
Nevertheless, they can still produce enhancements in the scattering amplitude. 
Resonance poles are also found in the lower half-plane, but above the line $\Re{p} + \Im{p} = 0$, and are associated with unstable states. 
Poles below this line but with a non-zero imaginary part are known as virtual state poles with non-vanishing width~\cite{Ikeda:2011dx}. 
Like ordinary virtual poles, these are not associated with physical states but can still influence the amplitude near threshold. 
Moreover, the analytic structure of the $S$-matrix, specifically the Schwarz reflection principle, requires that poles with non-zero imaginary parts appear in complex-conjugate pairs.

In the multi-channel case, the number of Riemann sheets increases as $2^n$, where $n$ is the number of channels. 
These Riemann sheets are distinguished by the signs of the imaginary parts of the channel momenta. The physical sheet, or first Riemann sheet, is defined as the region of the complex energy plane where 
$\Im{p_i} > 0$ for
all channels $i$.
Other sign combinations of $\Im{p_i}$ are generally referred to as the unphysical sheet. In the case of two-channel scattering, there are three such unphysical sheets. The first is the $[bt]$ sheet, also known as the second Riemann sheet, which is directly connected to the physical scattering region between the two thresholds. Here, ''b" indicates that channel 1 is analytically continued to the lower half of its momentum plane $(\Im{p_1}<0)$ while “t” means that channel 2 remains on the upper half $(\Im{p_2}>0)$. The next is the $[bb]$ sheet, or third Riemann sheet, which is connected with the scattering region beyond the second threshold; in this case, both $\Im{p_1}$ and $\Im{p_2}$ are negative. Finally, the $[tb]$ sheet, or the fourth Riemann sheet, which is the farthest from the physical scattering region and corresponds to $\Im{p_1}>0$ and $\Im{p_2}<0$. The distinction among these sheets arises from how the analytic continuation is performed across the branch cuts introduced by the channel thresholds, and each sheet plays a crucial role in understanding the analytic structure of the scattering amplitude.

\subsection{Uniformization}
In two-channel scattering, the complex momentum plane for each channel exhibits branch cuts induced by the presence of the other channel. 
To simplify the analytic structure and remove these cuts, we introduce a uniformizing variable $\omega$ defined as
\begin{equation}
    \omega = \dfrac{q_1 + q_2}{\sqrt{\epsilon_2^2 - \epsilon_1^2}}
    \quad\text{and}\quad
    \dfrac{1}{\omega} = \dfrac{q_1 - q_2}{\sqrt{\epsilon_2^2 - \epsilon_1^2}}.
    \label{eq:omega}
\end{equation}
This transformation maps the four-sheeted energy surface of the two-channel problem onto a single complex $\omega$-plane, allowing all Riemann sheets to be visualized within a unified representation, as shown in Fig.~\ref{fig:ambiguity}a.

The effect of a pole confined to a specific unphysical sheet can already be inferred from this mapping. 
For instance, a single pole in the lower half of the $[bb]$ sheet produces an enhancement only above the second threshold, while a pole in the upper half of the $[tb]$ sheet gives rise to a cusp precisely at the second threshold. 
When multiple poles are present across different sheets, the resulting interference patterns and threshold behavior become significantly more complex, complicating the task of identifying the underlying line shape structure.

\begin{figure}[t!]
    \centering
    \includegraphics[width=\linewidth]{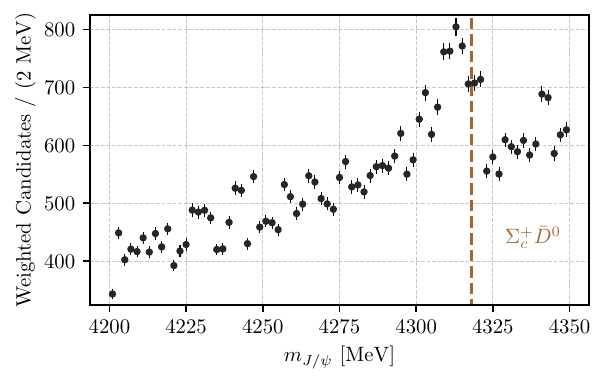}
    \caption{The relevant region of the invariant mass distribution of $J/\psi\,p$ from Ref.~\cite{lhcbcollaborationObservationNarrowPentaquark2019}. The vertical dashed line marks the $\Sigma_c^+\bar{D}^0$ threshold.}
    \label{fig:lhcb}
\end{figure}

\subsection{Independent \texorpdfstring{$S$}{S}-Matrix Poles}

Constructing a balanced training dataset of line shapes requires precise control over the placement of poles across various unphysical Riemann sheets. 
Moreover, the pole configurations must remain independent of any specific pole trajectories, which are typically tied to particular dynamical models. 
To ensure physical consistency, the parameterization must satisfy the minimal requirements of unitarity, analyticity, and correct threshold behavior. 
These conditions can be fulfilled by directly expressing the $S$-matrix in terms of independently specified Jost-like functions~\cite{rakityansky2023jost,santosInterpretationThresholdPeaks2023}.

We follow the formulation in Refs.~\cite{katoAnalyticalPropertiesTwochannel1965,santosInterpretationThresholdPeaks2023} to directly parameterize the $S$-matrix. 
The Jost-like function for our two-channel system takes the form
\begin{align}
    & D_m(q_1, q_2) = D_m(\omega) \\
    &= \dfrac{1}{\omega^2}
    (\omega - \omega_m)
    (\omega + \omega_m^*)
    (\omega - \omega_{m'})
    (\omega + \omega_{m'}^*),
    \label{eq:jost}
\end{align}
where $\omega_m$ denotes the physical pole, which can be freely assigned, and $\omega_{m'}$ is a regulator pole chosen to satisfy $|\omega_m \omega_{m'}| = 1$. 
We designate $\omega_m$ as the relevant pole by placing it closer to the scattering region, while $\omega_{m'}$ is assigned a phase of $e^{-i\pi/2}$ to minimize its impact on the line shape.

The structure of an unstable hadronic state is encoded in the distribution of poles across Riemann sheets~\cite{morganPoleCountingResonance1992,morganF0SMoleculeQuark1991,morganDecay$fracJensuremathpsiensuremathrightarrowensuremathvarphimathrmMM$Demands1993}. 
To account for various possible configurations, we specify the desired number of poles and construct the full Jost-like function as
\begin{equation}
    D(q_1, q_2) = \prod_{m=1}^{M} D_m(q_1, q_2),
    \label{eq:fulljost}
\end{equation}
where $M$ is the total number of poles. 
The elements of the two-channel $S$-matrix are then given by
\begin{equation}
\begin{split}
    S_{11} &= \dfrac{D(-q_1, q_2)}{D(q_1, q_2)},
    \quad
    S_{22} = \dfrac{D(q_1, -q_2)}{D(q_1, q_2)}, \\
    \det S &= \dfrac{D(-q_1, -q_2)}{D(q_1, q_2)}.
    \label{eq:smat}
\end{split}
\end{equation}
The relevant observable is the scattering amplitude, given as a linear combination of $T$-matrix elements, with $S = 1 + 2iT$. 
The structure of Eq.~\eqref{eq:smat} ensures unitarity of the $S$-matrix and guarantees hermiticity below the lowest threshold~\cite{newtonStructureManyChannelMatrix1961,couteurStructureNonrelativisticSmatrix1997}. 
Analyticity is also preserved, since pole types can be independently specified without violating the analytic structure.

The freedom to assign $S$-matrix poles independently allows us to expose the inherent ambiguity in interpreting line shapes. 
As illustrated in Fig.~\ref{fig:ambiguity}, two distinct pole configurations may result in identical near-threshold enhancements. 
This ambiguity stems from the cancellation of poles in one of the diagonal elements of the $S$-matrix. 
For example, a pole in the $[bb]$ sheet can cancel the effect of a pole in the $[tb]$ sheet, effectively removing their combined contribution from the $S_{11}$ element. 
Nevertheless, these poles still influence the $S_{22}$ element, allowing for a potential distinction between configurations. 
In practice, however, access to such complementary $S$-matrix elements requires additional experimental data from other processes, which is often unavailable in practice.

\subsection{Problem Setup: Analyzing the \texorpdfstring{$P_{c\bar{c}}(4312)^+$}{Pc(4312)+}}

In this manuscript, the end goal is to examine the conflicting interpretations of the $P_{c\bar{c}}(4312)^+$ signal. 
This candidate hadronic resonance is the lowest-mass hidden-charm pentaquark state, observed as an enhancement in the $J/\psi\,p$ invariant mass spectrum from the three-body decay $\Lambda_b^0 \rightarrow J/\psi\, p\, K^{-}$~\cite{collaborationObservation$JPsp$2015,lhcbcollaborationObservationNarrowPentaquark2019}. 
The enhancement appears near the $\Sigma_c^+\bar{D}^0$ threshold, prompting numerous studies to propose a meson-baryon molecular interpretation~\cite{guoHadronicMolecules2018,xiaoExploringMolecularScenario2019,duInterpretationLHCb$P_c$2020}. 
However, alternative scenarios remain viable: 
some propose a compact pentaquark state, while others interpret it as a virtual state tied to the $\Sigma_c^+\bar{D}^0$ channel.

We restrict our analysis to the narrow energy window around the $\Sigma_c^+\bar{D}^0$ threshold, from $4200$ GeV to $4350$ GeV. This choice ensures that the extracted pole structure corresponds to the enhancement region with well-defined significance, as shown in Fig.~\ref{fig:lhcb}. Furthermore, we further assume that the nature of the enhancement is due to poles near the scattering region since the triangle singularity interpretation is already ruled out~\cite{lhcbcollaborationObservationNarrowPentaquark2019,Co:2024bfl}.

\begin{figure}[t!]
    \centering
    \includegraphics[width=0.75\linewidth]{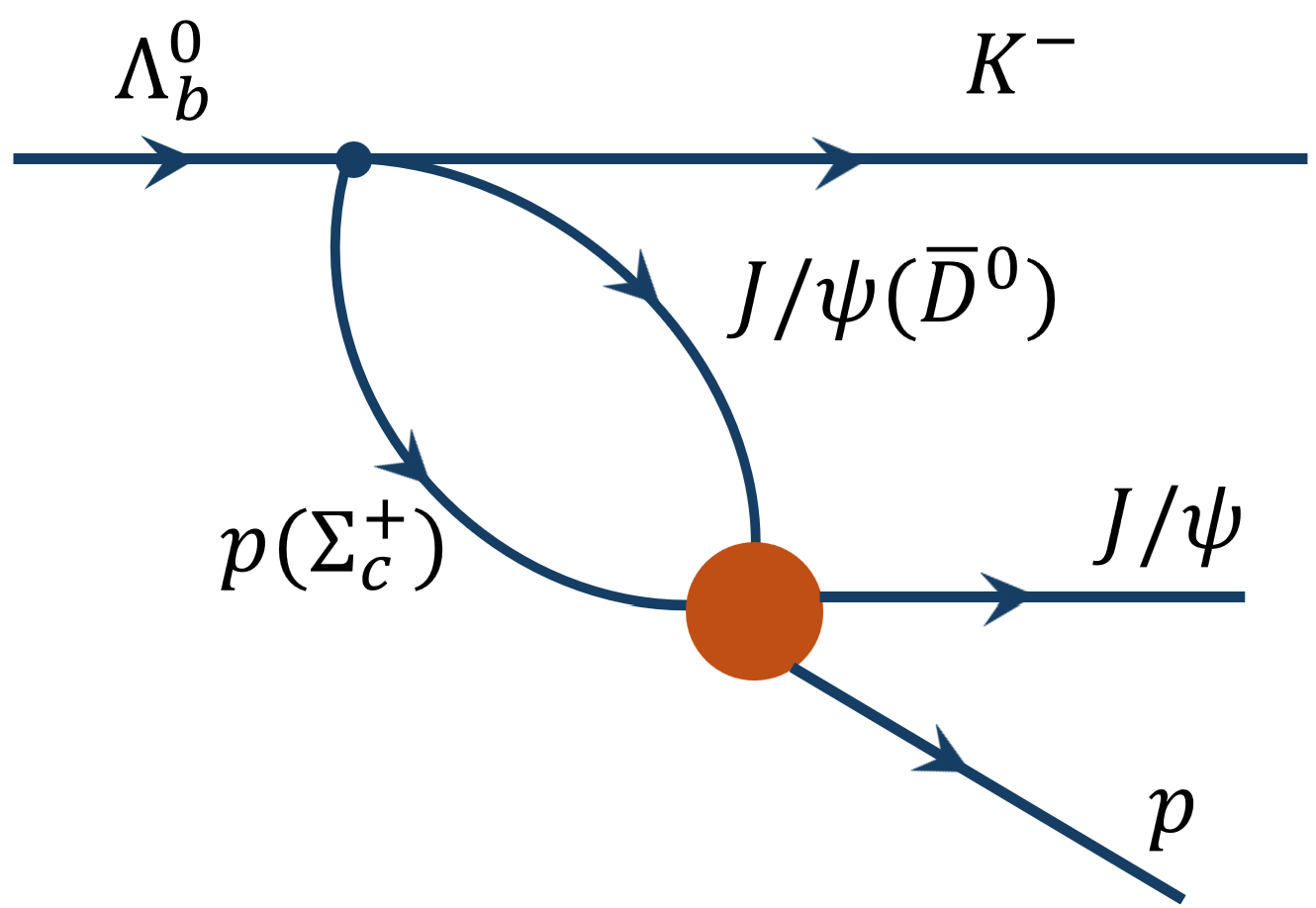}
    \caption{The reaction process for the decay $\Lambda_b^0 \rightarrow J/\psi\,p\,K^-$.} 
    \label{fig:feynmann}
\end{figure}

We model the reaction using the diagram in Fig.~\ref{fig:feynmann}. 
The vertex directly connecting $\Lambda_b^0$ to the three-body intermediate state is parametrized by the three-body phase space and a given production rate constant. 
The important part of the analysis is encoded in the two-body interaction connecting the initial intermediate channel to the final state. 
Using these considerations, we use the fitting function
\begin{equation}
    \dfrac{dN}{d\sqrt{s}} = \rho\left[|F(s)|^2 + B(s)\right],
    \label{eq:fitfun}
\end{equation}
where $\rho$ is a phase space factor and $B(s)$ encodes the smooth background contribution, which will suffice for our present purpose. We parametrize the background $B(s)$ using a sixth-degree polynomial $B(s)=\sum_{n=0}^6\alpha_n\left(\sqrt{s}\right)^n$ where $\alpha_n\in(0,3.25)$ for $n=0,1,\cdots 6$.

The signal term is modeled as $F(s) = \sum_{i=1}^2 \beta_i T_{1i}(s)$, where $\beta_i$ denotes the production strength of channel $i$ sampled in the interval $\beta_i\in(0.1,1.5)$, and $T_{1i}$ is an element of the $T$-matrix:
\begin{equation}
T =
\begin{pmatrix}
T_{11} & T_{12} \\
T_{21} & T_{22}
\end{pmatrix}
=
\begin{pmatrix}
T_{J/\psi\,p \rightarrow J/\psi\,p} & T_{J/\psi\,p \rightarrow \Sigma_c\bar{D}} \\
T_{\Sigma_c\bar{D} \rightarrow J/\psi\,p} & T_{\Sigma_c\bar{D} \rightarrow \Sigma_c\bar{D}}
\end{pmatrix}.
\label{eq:Tmatrix}
\end{equation}

Synthetic line shapes used for training are generated from Eq.~\eqref{eq:fitfun}, with pole structures defined via the assigned singularities of the $T$-matrix. 
For this study, we consider configurations containing up to four poles distributed across the unphysical Riemann sheets: $[bt]$, $[bb]$, and $[tb]$. The four-pole restriction is enough to accommodate the possibility of two genuine resonances which requires two pole-shadow pairs. 
To ensure systematic coverage, poles are sampled within the relevant region of each sheet, bounded by proximity to the $\Sigma_c^+\bar{D}^0$ threshold and constrained to lie within a narrow complex energy strip around the physical axis. That is,
\begin{equation}
    \begin{cases}
        T_2-50 \leq 
        \Re{E}_\text{pole}\leq 4350\; & \text{all RS}\\
        -100 \leq \Im{E}_\text{pole} < 0 & \text{[$bt$] \& [$bb$]}\\
        0 < \Im{E}_\text{pole} \leq 100 & \text{[$tb$]}.
    \end{cases}
    \label{eq:region}
\end{equation}
This sampling strategy ensures that the generated line shapes reflect physically plausible near-threshold dynamics without relying on specific pole trajectories.

\section{Algorithm}
\label{sec:algorithm}
This section presents a machine learning framework for inferring pole configurations from data, along with estimates of predictive uncertainty.
The primary goals are to construct effective input representations from the generated data, identify suitable model architectures for the learning task, and establish methods for estimating and interpreting uncertainty in the model’s predictions.
To that end, we describe the structure of the training data, the extraction of task-relevant features, and the application of Gradient Boosting models tailored to this problem.
We also explore different model variants and discuss how uncertainty-aware inference can be used to identify unreliable predictions and guide model interpretation.

\subsection{Dataset}
\label{sec:dataset}
The dataset $\mathcal{X}$ contains $3.5 \times 10^{5}$ individual measurements, referred to as raw features in Fig.~\ref{fig:overview}a.
Each data point $\boldsymbol{x} \in \mathcal{X}$ is a real-valued vector in $\mathbb{R}^{75}$, where the $75$ dimensions correspond to evaluations of a synthetic line shape at discrete energy values.
More precisely, each feature represents the value of the distribution $\frac{dN}{d\sqrt{s}}$ at a specific energy point, as defined in Eq.~\eqref{eq:fitfun}.
This distribution incorporates contributions from the coupled-channel $T$-matrix elements via the function $F(s)$ and captures relevant physical phenomena such as resonance peaks, interference effects, and threshold behavior arising from different pole configurations.
The samples span $35$ distinct pole configurations, each corresponding to a unique combination of pole counts in the $[bt]$, $[bb]$, and $[tb]$ regions, which together define the dataset labels $\boldsymbol{y}$.
For example, a configuration labeled as $([3], [1], [0])$ corresponds to three poles placed on the $[bt]$ sheet, one pole on the $[bb]$ sheet, and none on the $[tb]$ sheet.
These configurations include all non-negative integer combinations such that the total number of poles satisfies $[bt] + [bb] + [tb] \leq 4$, with each region containing between $0$ and $4$ poles.

\subsection{Feature Extraction}
\label{sec:features}

Machine learning models benefit from input representations that are both mathematically and computationally convenient to process~\cite{lecunDeepLearning2015,Goodfellow-et-al-2016,princeUnderstandingDeepLearning2023}.
In general, there are two complementary approaches to obtaining such representations: representation learning and feature engineering.
Representation learning encompasses techniques that enable models to automatically infer useful representations from raw data, tailoring them to the task at hand through training~\cite{bengioRepresentationLearningReview2013}.
Feature engineering involves the deliberate transformation of raw data into more informative and structured inputs, typically guided by domain expertise~\cite{zhengFeatureEngineeringMachine2018,dongFeatureEngineeringMachine2018}.
Both approaches share a common goal: to expose relevant information to the machine learning model in order to improve predictive performance.

In recent years, representation learning, particularly through deep neural networks, has become the dominant paradigm in domains such as computer vision~\cite{krizhevskyImageNetClassificationDeep2012,heDeepResidualLearning2016} and natural language processing~\cite{devlinBERTPretrainingDeep2019,vaswaniAttentionAllYou2017}.
This trend is largely motivated by the challenges of handcrafting generalizable features for complex, high-dimensional data such as images, videos, and natural language.
Modern deep learning architectures can effectively discover hierarchical representations directly from raw inputs, reducing or eliminating the need for manual feature design.

Nonetheless, feature extraction remains highly relevant in settings where domain-specific insights can guide the identification of informative structures in the data; especially when working with structured or tabular formats~\cite{kuhnFeatureEngineeringSelection2019, jannachWhyAreDeep2020}.
In such contexts, carefully engineered features can be both interpretable and enable accurate predictions.

In this study, we perform extensive feature engineering with the goal of capturing representations that effectively distinguish between different pole structures.
Our prior knowledge that the one-dimensional signals under study are sequential and ordered by an independent variable (energy) naturally motivates the use of techniques from time series analysis~\cite{christTimeSeriesFeatuRe2018,lubbaCatch22CAnonicalTimeseries2019,fulcherHighlyComparativeFeaturebased2014,bagnallGreatTimeSeries2017}.

As illustrated in Fig.~\ref{fig:overview}a, we begin by extracting a comprehensive set of time series features from the raw line shapes and their associated energy values.
To reduce dimensionality and retain only the most informative features, we then apply statistical selection methods, following the procedure described in Ref.~\cite{christTimeSeriesFeatuRe2018}.
The selected features can be grouped into several categories: statistical descriptors, temporal descriptors, frequency-domain characteristics, distribution descriptors, entropy-based measures, and (non-)linear correlations.

\begin{figure}[t!]
    \centering
 \includegraphics[width=\linewidth]{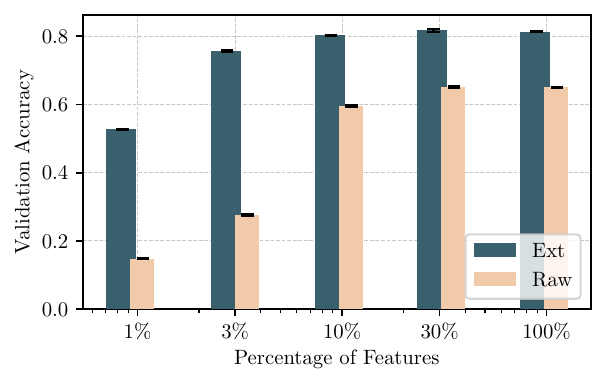}
    \caption{\textbf{Feature Selection:}  
    Performance comparison between models trained on raw features and those trained on extracted features.
    A baseline model is first trained on the full dataset to rank feature importance.
    Subsequent models, whose performance is shown here, are independently trained on subsets containing varying percentages of the top-ranked features.
    The results show that extracted features consistently outperform raw features, with diminishing performance gains beyond the top $30\%$ of features.
    Results are averaged using 5-fold cross-validation; error bars represent the standard deviation across folds.}\label{fig:feature-selection}
\end{figure}

To evaluate the impact of feature selection on model performance, we compare predictive models trained on raw input features with those using extracted features.
As a baseline, we first train a Gradient Boosting classifier on the complete respective dataset to obtain a ranked list of feature importances derived from the model~\cite{natekinGradientBoostingMachines2013}.
These importance scores are based on the total gain accumulated across all tree splits where a feature is used, allowing us to rank features by their predictive utility.
In tree-based models like Gradient Boosting, each split partitions the data to reduce the prediction error, and features that contribute more to this error reduction are considered more important.
This approach effectively quantifies how useful a feature is in improving accuracy during model training, making it a useful tool for feature selection.
We then train additional models on subsets of the most important features, retaining varying percentages of the top-ranked features.
This procedure allows us to assess how model accuracy depends on the number of features used and highlights which types of information contribute most substantially to the learning task.
As shown in Fig.~\ref{fig:feature-selection}, models trained on extracted features consistently outperform those trained on raw line shapes.
Moreover, we observe diminishing returns when retaining more than the top $30\%$ of features, indicating that most of the predictive signal is captured by a relatively small subset.
All results are averaged using 5-fold cross-validation, where the dataset is split into five parts, and each part is used once as a validation set while the remaining four are used for training. 
This provides a robust estimate of model performance across different data partitions, with error bars reflecting the standard deviation across folds.
Finally, this approach allows us to select the top $30\%$ of features as the final subset, which rounds up to the top $128$ extracted distinct features comprising the modified data corpus $\tilde{\mathcal{X}}$.
The complete list of extracted features with their importance scores is provided together with the source code in the accompanying repository~\cite{frohnert2025polelearning}. 

\subsection{Model Selection}
\label{sec:model-selection}
The tabular features extracted in Sec.~\ref{sec:features} naturally motivate the use of Gradient Boosting methods, which are regarded as the state of the art for tabular data~\cite{natekinGradientBoostingMachines2013,borisovDeepNeuralNetworks2024,shwartz-zivTabularDataDeep2021}.
In this manuscript, we use CatBoost, which offers a particularly efficient and robust implementation of Gradient Boosting with decision trees~\cite{dorogushCatBoostGradientBoosting2018,prokhorenkovaCatBoostUnbiasedBoosting2019}.

In general, Gradient Boosting builds an ensemble of weak learners, typically decision trees, by iteratively training each new model to correct the residual errors of the previous ones.
At every step, the algorithm fits a tree to the gradient of the loss function with respect to the current model prediction, progressively improving the overall accuracy.
A feature of tree-based models like CatBoost is their invariance to monotonic transformations of input features~\cite{prokhorenkovaCatBoostUnbiasedBoosting2019}, which allows us to directly train on the extracted features without additional preprocessing.
Training details of the Gradient Boosting model implementation are discussed in Appx.~\ref{appx:training}.

\subsection{Designing the Learning Task}
\label{sec:class_vs_reg}

A central challenge in this work is the formulation of the learning task for the machine learning model.
In the most general setting, the goal is to learn a function $f:\tilde{\mathcal{X}} \mapsto \mathcal{Y}$ that maps input data $x \in \tilde{\mathcal{X}}$ to corresponding target labels $y \in \mathcal{Y}$.
The learning process involves minimizing a loss function $L=\sum \ell(y, f(x))$, which quantifies the sum of discrepancies between predicted and true outputs over the training dataset.
The structure of the label space $\mathcal{Y}$ and the choice of the loss function $L$ play a crucial role in shaping the learning dynamics and influencing generalization~\cite{princeUnderstandingDeepLearning2023}.
We consider two distinct formulations of the learning problem: 
(1) multi-class learning over the complete pole structures simultaneously, and (2) a decomposed approach where the learning task is separated by pole type.
The motivations behind the two formulations are two-fold: 
first, the structure and dimensionality of the target labels directly shape the training dynamics; 
second, the per-pole-type prediction approach, unlike the per-class formulation, yields more detailed insights into the model’s behavior, which we expect to provide a clearer understanding of how the underlying physics manifests in the learning task.

In general, the learning task is to predict the non-negative integer pole counts $k_i \in \mathbb{Z}_{\geq 0}$ that satisfy  
\begin{equation}
    \sum_{i=1}^{s} k_i \leq n,
\end{equation}
where $s$ denotes the number of distinct Riemann sheets, $k_i$ is the number of poles in the $i^{th}$ Riemann sheet, and $n$ specifies the total number of poles.

In the multiclass formulation, which has been studied in previous works~\cite{sombilloClassifyingPoleAmplitude2020,sombilloClassifyingThresholdEnhancement2021,sombilloModelIndependentAnalysis2021}, the label space is defined as $\boldsymbol{y} \in \{0,1\}^{K(n, s)}$, where  
\begin{align}
K(n, s) = \binom{n + s}{s}.
\end{align}  
This formulation treats each valid pole configuration as a distinct class. 
In the setting considered in this manuscript, with $s = 3$ and $n = 4$, valid configurations correspond to all non-negative integer tuples $(k_1, k_2, k_3)$ satisfying $k_1 + k_2 + k_3 \leq 4$ (e.g., $([bt], [bb], [tb]) = (0, 0, 0)$, $(0, 0, 4)$, or $(1, 1, 2)$), each mapped to a unique class index and represented as a one-hot vector.
The size of the label space grows polynomially with $n$ for fixed $s$, and polynomially with $s$ for fixed $n$.

We propose an alternative formulation of the learning problem by decomposing the target by pole type, which yields a matrix $\boldsymbol{Y} \in \{0,1\}^{s \times (n+1)}$, where each row $i$ corresponds to sheet $i$ and contains a one-hot encoding of the pole count $k_i \in \{0, 1, \ldots, n\}$ for that sheet. 
This yields a structured output in which the information on pole counts for each sheet is encoded independently, rather than encoding the entire configurations jointly.
In the case of $s = 3$ and $n = 4$, this corresponds to predicting a $3 \times 5$ binary matrix, where each row is a one-hot vector specifying the number of poles (0 through 4) for sheets $[bt]$, $[tb]$, and $[bb]$, respectively. 
A key advantage of this formulation is that the label dimensionality scales linearly, in contrast to the combinatorial growth of the multiclass approach. 
This may lead to improved training efficiency and scalability, particularly in regimes where either $n$ or $s$ is large.
Moreover, from an explainability perspective, the decomposed formulation enables per-sheet prediction analysis, offering more granular insight into the model's behavior and decision process.
An additional regression-style labeling approach is discussed in Appx.~\ref{appx:regression}.

\begin{figure}[t!]
    \centering
 \includegraphics[width=\linewidth]{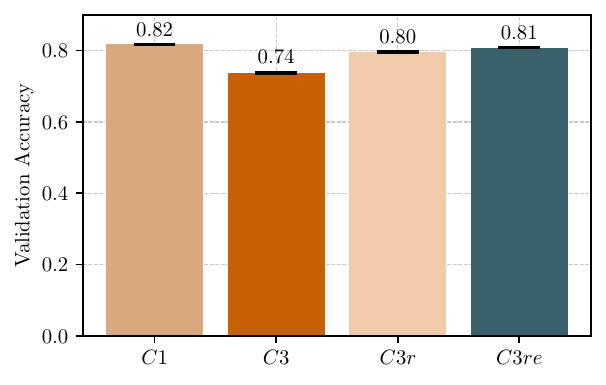}
    \caption[]{\textbf{Designing the Learning Task:} 
Five-fold cross-validation accuracy for various modeling strategies used to formulate the learning task.
The model labeled with $1$ is trained using directly discretized labels in the space \(\boldsymbol{y} \in \{0,1\}^{K(n)}\), whereas those labeled with $3$ use structured labels in $\boldsymbol{Y} \in \{0,1\}^{s \times (n+1)}$. 
$C1$ and $C3$ denote standard classification models with one and three output heads, respectively.
$C3r$ and $C3re$ are chained (recursive) models, with $C3re$ representing an ensemble of such chains.
The results compare the predictive performance of different classification formulations discussed in the text.
All models are trained to predict pole configurations across the $[bt]$, $[tb]$, and $[bb]$ regions, using either shared or factorized label representations accordingly.
    }\label{fig:clf-performance}
\end{figure}

However, this latter formulation introduces an additional challenge: constructing a machine learning model which predicts matrices $\boldsymbol{Y} \in \{0,1\}^{s \times (n+1)}$, where each row corresponds to a probability distribution over the possible pole counts at a given position. 
In this manuscript, we make the choice to decouple the rows of $\boldsymbol{Y}$ such that each label encodes the predicted number of poles at the respective sheet: $\boldsymbol{Y}_0: \boldsymbol{y}_{bt} \in \{0,1\}^{n+1}$, $\boldsymbol{Y}_1: \boldsymbol{y}_{bb} \in \{0,1\}^{n+1}$, and $\boldsymbol{Y}_2: \boldsymbol{y}_{tb} \in \{0,1\}^{n+1}$. 
Exploring additional modeling strategies is left for future research.

The following analysis examines how these two formulations influence the learning dynamics.
We evaluate the performance of the following models:
\begin{itemize}
    \item $C1$: A single model trained on the full discretized label space, $\boldsymbol{y} \in \{0,1\}^{K(n,s)}$, treating each pole configuration as a distinct class.
    \item $C3$: A decomposed approach using three separate models, each dedicated to one pole type. 
    The models are trained independently to predict $\boldsymbol{y}_{bt}$, $\boldsymbol{y}_{tb}$, and $\boldsymbol{y}_{bb}$, respectively.
    \item $C3r$: A sequential model chain consisting of three submodels, where each model predicts one pole type (${\boldsymbol{y}}_{bt}$, ${\boldsymbol{y}}_{tb}$, ${\boldsymbol{y}}_{bb}$) and passes its prediction as additional input to the next model in the chain. 
    The model chain is trained using the full structured target representation $\boldsymbol{Y} \in \{0,1\}^{s \times (n+1)}$, where each submodel learns the pole count distribution for a specific sheet. 
    Unlike the $C3$ baseline, the submodels in this chain are conditionally dependent, allowing the model to capture correlations between different pole types.
    \item $C3re$: Same as $C3r$, but using an ensemble of $M$ predictor chains. 
    In addition to enabling uncertainty estimation, ensembles are also known to improve predictive performance through variance reduction and model averaging.
\end{itemize}

To evaluate how different formulations of the learning task affect predictive performance, we compare four modeling strategies using five-fold cross-validation. 
The resulting validation accuracy is shown in Fig.~\ref{fig:clf-performance}.
Model $C1$, trained on directly discretized labels, achieves higher accuracy than model $C3$, which uses separate classifiers for each pole type. 
This gap in performance can be attributed to the fact that the $C3$ models are trained independently and do not share information, making it difficult to learn global structural constraints, such as $[bt] + [bb] + [tb] \leq 4$, that describe the complete pole configuration.
Model $C3r$ addresses this limitation by correlating the outputs through a recursive chain, effectively passing information between the submodels.
The order of the chain is chosen based on the validation accuracies achieved by the $C3$ model: 
since not only additional information but also prediction errors propagate through the chain, we place the label with the highest accuracy, $[bt]$, first, followed by $[bb]$, and finally the most challenging, $[tb]$.
Finally, $C3re$, an ensemble of such chained predictors, achieves performance comparable to $C1$ while benefiting from a more scalable label space. 
Moreover, $C3re$ retains the explainability of per-pole-type predictions, which is crucial for estimating uncertainty in the next section.

\subsection{Uncertainty Estimation}
\label{sec:uncertainty}
\begin{figure}[t!]
    \centering
 \includegraphics[width=0.84\linewidth]{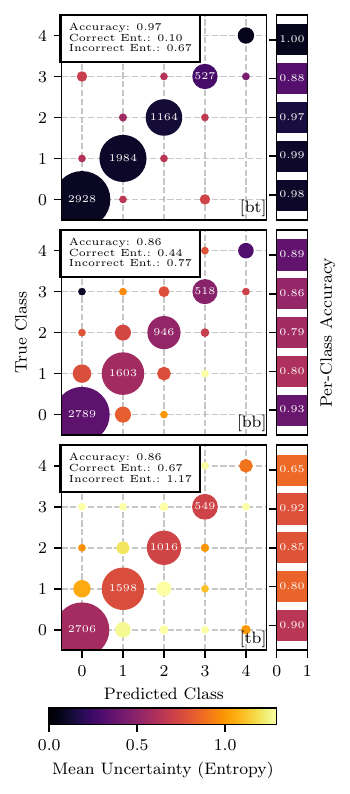}
    \caption[]{\textbf{Estimating the Uncertainty:}
Visualization of classification performance and associated uncertainty for the three pole types: $[bt]$, $[bb]$, and $[tb]$.
Each panel displays a confusion matrix, where the position of each circle encodes predicted vs. true number of poles in the sheet, circle size reflects the number of samples in that category, and color indicates the cumulative mean predictive uncertainty (entropy).
Darker colors correspond to lower uncertainty.
Horizontal bars to the right of each panel show per-class accuracies, with numerical values annotated.
Summary statistics in the top-left corner of each subplot report the overall accuracy, along with separate average predictive entropy values for correctly and incorrectly predicted samples.
This representation highlights not only the distribution of model errors but also their associated uncertainty, revealing systematic differences in model confidence across prediction types and pole types. 
    }\label{fig:uncer}
\end{figure}

In many machine learning applications, particularly those with a low tolerance for error, such as medical diagnostics or autonomous driving, it is not sufficient to rely solely on the model’s prediction for decision making~\cite{amodeiConcreteProblemsAI2016,guoCalibrationModernNeural2017,kendallWhatUncertaintiesWe2017}. 
Equally important is an assessment of how confident the model is in its output~\cite{galUncertaintyDeepLearning2016,abdarReviewUncertaintyQuantification2021}. 
This consideration is also central to our work, where incorrect inferences about the underlying pole structure could lead to misleading or unjustified conclusions about the physical interpretation.

To address this, we rely on predictive uncertainty estimation; a framework that quantifies the confidence associated with model outputs~\cite{lakshminarayananSimpleScalablePredictive2017}. 
Uncertainty is typically divided into two main categories: aleatoric and epistemic~\cite{kendallWhatUncertaintiesWe2017}.
Aleatoric uncertainty arises from inherent noise or ambiguity in the data, such as overlapping classes. 
It is a property of the data distribution itself and cannot be reduced even by collecting more data or training a better model~\cite{depewegDecompositionUncertaintyBayesian2017}.
Epistemic uncertainty, in contrast, reflects the model's lack of knowledge, for instance due to insufficient or non-representative training data. 
It captures uncertainty over the model parameters and is high in regions of the input space not well covered by the training distribution. 
Unlike aleatoric uncertainty, it can be reduced by gathering additional data in those regions~\cite{depewegDecompositionUncertaintyBayesian2017}.

To jointly quantify both sources of uncertainty, we employ an ensemble of models, each trained independently on the same dataset. 
In classification tasks, the resulting class probability vectors can be analyzed using entropy-based measures, which provide a natural quantification of predictive uncertainty~\cite{malininUncertaintyGradientBoosting2021}.

Let $\{f_i(\boldsymbol{x})\}_{i=1}^M$ be an ensemble of $M$ trained models making predictions on an input $\boldsymbol{x}$.
Each model $f_i$ outputs a class probability vector $p_i = f_i(\boldsymbol{x})$, where $p_i \in [0,1]^C$ and $\sum_{c=1}^C p_{i,c} = 1$.
Here, $C$ denotes the number of possible classes, and $c$ indexes the class dimension within each probability vector.
The ensemble thus produces a set of $M$ such class probability vectors: $\{p_i\}_{i=1}^M$, where $i$ indexes the models in the ensemble.
The total predictive uncertainty is then computed as the entropy of the ensemble-averaged prediction~\cite{malininUncertaintyGradientBoosting2021}:

\begin{align}
\text{Total Uncertainty} = H(\bar{p}) = -\sum_{c=1}^C \bar{p}_c \log \bar{p}_c, \label{eq:uncertainty}
\end{align}

where $\bar{p} = \frac{1}{M} \sum_{i=1}^M p_i$ is the mean class probability vector across the ensemble.
Intuitively, if the individual models $f_i$ strongly disagree (i.e., assign high probability mass to different classes) the averaged distribution $\bar{p}$ becomes more uniform, leading to a higher entropy $H(\bar{p})$, and thus a higher total uncertainty.
The aleatoric uncertainty is estimated as the mean entropy of the individual model predictions:

\begin{align}
\text{Aleatoric Uncertainty} &= \frac{1}{M} \sum_{i=1}^M H(p_i),
\end{align}

where $H(p_i)=-\sum_{c=1}^C p_{i,c} \log (p_{i,c})$ is the entropy of the class probability vector from model $i$.
The epistemic uncertainty is then obtained as the difference between the total and aleatoric uncertainty:

\begin{align}
\text{Epistemic Uncertainty} &= H(\bar{p}) - \frac{1}{M} \sum_{i=1}^M H(p_i).
\end{align}
This decomposition allows us to distinguish uncertainty arising from inherent data ambiguity (aleatoric) from that due to limited model knowledge or variability across the ensemble (epistemic).

Using the ensemble model $C3re$ introduced in Sec.~\ref{sec:class_vs_reg}, we extract the (total) predictive uncertainty estimates for each individual pole.
We emphasize that such per-pole-type uncertainty quantification is only possible with $C3$-style models. 
In contrast, $C1$-style models yield only per-class uncertainties, making it impossible to disentangle which specific pole contributes to the uncertainty in a given prediction.
The results are presented in Fig.~\ref{fig:uncer}.
Each panel of the figure corresponds to one of the three pole regions ($[bt]$, $[bb]$, and $[tb]$) and visualizes classification performance alongside uncertainty. 
In each panel, the confusion matrix is shown using circles:
the position encodes the predicted vs. true number of poles, the size reflects the number of samples in that bin, and the color encodes the cumulative mean predictive uncertainty (measured via entropy). 
Darker colors indicate lower uncertainty. 
In this context, cumulative refers to the need to account for the correlation between uncertainties and the sequential nature of predictions in a classifier chain. 
The first model in the chain, $f_{[bt]}$, makes a prediction that is passed to the next model, $f_{[bb]}$, and so on. 
However, during training, these models do not have access to the predictive uncertainty of previous stages, as uncertainty estimates are only available at inference time through posterior sampling. 
To obtain a consistent notion of uncertainty across the chain, we define cumulative uncertainty as follows: 
$\bar{ \mathcal{H}}_{[bt]} \coloneqq \mathcal{H}_{[bt]}$, $\bar{ \mathcal{H}}_{[bb]}\coloneqq\mathcal{H}_{[bt]}+\mathcal{H}_{[bb]}$, and $\bar{ \mathcal{H}}_{[tb]}\coloneqq\mathcal{H}_{[bt]}+\mathcal{H}_{[bb]}+\mathcal{H}_{[tb]}$.
This approach allows us to analyze how uncertainty propagates through the chain and influences downstream predictions and is motivated in Appx.~\ref{appx:cascading}.
On the right of each panel, horizontal bars indicate the per-class accuracy, with the color encoding the cumulative average predictive uncertainty for each class. 
In the top-left corner of each subplot, summary statistics report the overall accuracy and the average uncertainty, broken down into correct and incorrect predictions.
A perfect classifier would result in diagonal matrices with dark spheres.

We note several key observations:
The model performs best on $[bt]$ predictions ($97\%$ accuracy), followed by $[bb]$ ($86\%$), and $[tb]$ ($86\%$). 
This trend likely reflects intrinsic differences in data ambiguity between the regions. 
While both the $[bt]$ and $[bb]$ sheets are relatively close to the physical sheet, the model’s superior performance on $[bt]$ can be attributed to the fact that more data points in the chosen energy region are directly connected to the $[bt]$ sheet, whereas significantly fewer points are connected to the $[bb]$ sheet. 
This imbalance in direct connectivity likely accounts for the observed hierarchy. 
Although the $[tb]$ sheet can only produce a structure exactly at the threshold, its poles still influence the amplitude in the nearby energy region, both below and above the threshold, allowing the model to make use of surrounding points. 
This likely explains why $[tb]$ accuracy is comparable to $[bb]$, despite the $[tb]$ sheet’s structural constraints.

The largest misclassification errors occur predominantly between neighboring classes, suggesting a form of local ambiguity in label space.
For example, in the $[tb]$ region (Fig.~\ref{fig:uncer}), a common error is predicting class 0 when the true label is 1; indicating overlap in feature space between adjacent classes.
As the distance between true and predicted classes increases, misclassifications become less frequent.
Notably, the model rarely confuses class 0 with class 4.
This trend is visually reflected in the decreasing size of off-diagonal circles and the overall drop-off in error frequency with increasing class distance from the diagonal.

Notably, predictive uncertainty correlates strongly with prediction correctness.
In the $[bt]$ region, predictions for class 0 are correct in $98\%$ of cases, and the associated mean predictive entropy is close to zero (visualized in black), indicating that the ensemble’s mean probability vector is sharply peaked on a single class.
Conversely, higher uncertainty is associated with misclassifications.
In the $[bb]$ region, for instance, off-diagonal elements, corresponding to false predictions, appear progressively lighter in color as one moves away from the diagonal.
This indicates that larger classification errors are accompanied by higher predictive uncertainty.
Also on the diagonal, classes with lower per-class accuracy tend to appear lighter, reflecting the model’s uncertainty in more ambiguous cases.
For example, a mean entropy of around 0.5 for predicted class 1 in the $[bb]$ sheet indicates that the model favors class 1, but still assigns appreciable probability mass to neighboring classes such as 0 and 2.
This distribution reflects mild disagreement within the ensemble or local ambiguity in the feature space surrounding class 1.
In the $[tb]$ region, some predictions exhibit mean uncertainty values exceeding 1, indicating cases where the mean probability vector is relatively uniform and/or the previous predictions were very uncertain.
These typically arise when multiple stages of the classifier chain are both uncertain and incorrect, leading to compound errors and a near-uniform predictive distribution.

Finally, the summary statistics reveal a consistent gap in average entropy between correct and incorrect predictions across all regions.
In practice, this means that incorrect predictions tend to have more uniform (spread-out) probability distributions, while correct predictions tend to be more sharply peaked around a single class.
As a result, computing the entropy of a model’s output probability vector provides a reliable indication of its confidence: 
lower entropy corresponds to higher certainty, and vice versa.
These findings support the conclusion that the model’s uncertainty estimates are meaningful and systematically reflect its prediction confidence.
We note that our analysis focuses on the total predictive uncertainty and does not separately visualize the aleatoric and epistemic components. 
This choice is justified by the fact that we restrict ourselves to in-distribution data, where the data-dependent (aleatoric) uncertainty is minimal and does not warrant separate investigation. 
However, in scenarios involving out-of-distribution samples, such as test cases with a pole count not seen during training, e.g., a configuration with six total poles, one could leverage the data uncertainty to identify such deviations from the training distribution.

Overall, the visualization in Fig.~\ref{fig:uncer} not only captures the per-pole-type predictions of the model, but also how confident it is in those predictions; exposing structured patterns in both error frequency and uncertainty across pole regions and class labels.

\subsection{Rejecting Uncertain Predictions}
\label{sec:reject}
\begin{figure}[t!]
    \centering
 \includegraphics[width=\linewidth]{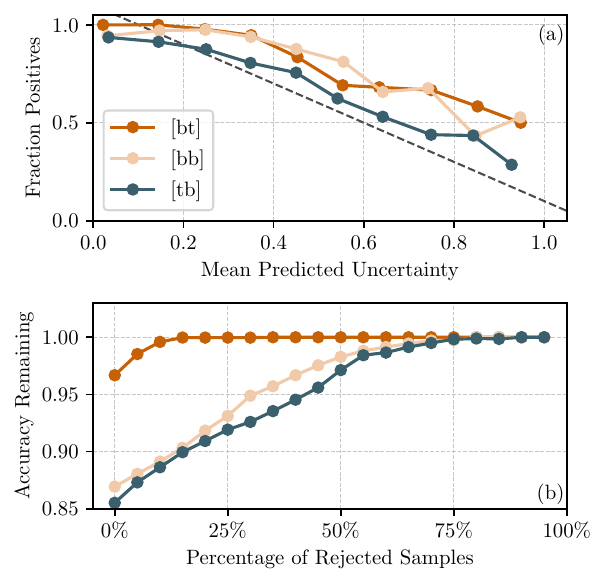}
    \caption[]{\textbf{Prediction-Rejection Ratio.}
The $C3re$ ensemble model is trained to predict the pole structures $[bt]$, $[tb]$, and $[bb]$.
The figure shows two complementary visualizations for evaluating predictive uncertainty.
(a) Calibration plot showing the fraction of correct predictions as a function of the predicted uncertainty $H(\bar{p})$.
All three curves follow the linear trend on the diagonal, meaning that the the uncertainty estimates reflect actual accuracy, indicating well-calibrated classifier.
(b) Accuracy-rejection curve based on cumulative predictive entropy $H(\bar{p})$.
Accuracy is computed after discarding an increasing percentage of high-uncertainty samples.
As more uncertain predictions are rejected, accuracy of the retained predictions steadily improves, confirming that the model’s uncertainty estimates are linked to correctness and can serve as a reliable rejection criterion.
}\label{fig:reject}
\end{figure}

In the inference stage of machine learning models, it is (naturally) essential to avoid being misled by incorrect predictions~\cite{lakshminarayananSimpleScalablePredictive2017}.
To this end, having access to a meaningful estimate of the model’s uncertainty or confidence is valuable, as it enables one to identify predictions that are likely to be incorrect.
As shown in Sec.~\ref{sec:uncertainty}, the uncertainty values produced by the ensemble appear to correlate strongly with the correctness of the model’s output.
Building on this observation, we now investigate whether predictive uncertainty can be used as a rejection criterion: that is, whether one can systematically discard uncertain predictions to improve the overall reliability of the retained subset.

\begin{figure*}[ht!]
    \centering
 \includegraphics[width=\linewidth]{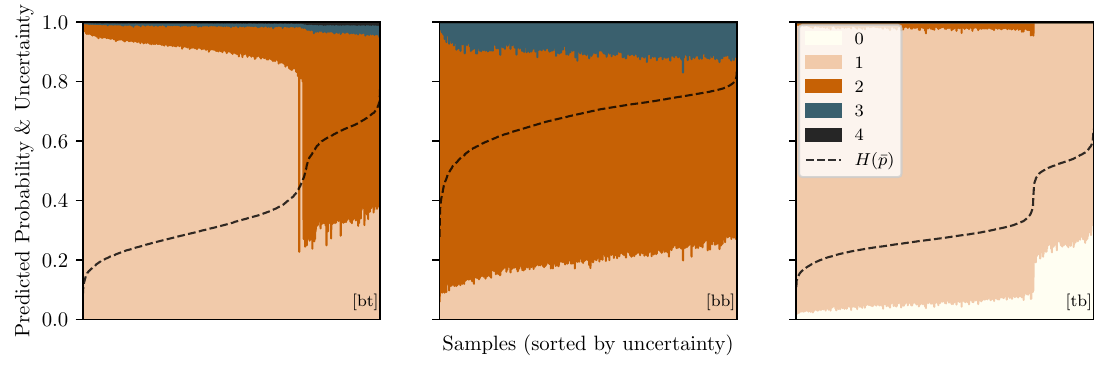}
    \caption[]{
    \textbf{Class Probabilities and Predictive Uncertainty for Experimental Data:} 
    Predicted class probabilities (stacked bar chart) and associated uncertainties (dashed line) for the pole structures $[bt]$, $[bb]$, and $[tb]$.
    Each panel shows the predicted class distribution for one target, sorted by increasing predictive entropy $H(\bar{p})$.
    Stacked color bands indicate the predicted class probabilities for each sample, while the dashed black line shows the corresponding uncertainty.
    Samples with high uncertainty exhibit more diffuse class distributions, whereas confident predictions are sharply peaked.
    The model expresses the highest overall certainty for $[bt]=[1]$, $[bb]=[2]$ and $[tb]=[1]$.
}\label{fig:exp}
\end{figure*}

We define a range of uncertainty (and confidence) thresholds and evaluate the model only on those samples whose predictive certainty exceeds each threshold.
If the uncertainty estimates are well-calibrated, then the subset of retained predictions, those with the lowest uncertainty, should be more accurate on average.
Conversely, discarding high-uncertainty (or low-confidence) predictions should improve the reliability of the model's output~\cite{lakshminarayananSimpleScalablePredictive2017}.

The $C3re$ ensemble model is trained to classify the presence of poles in the $[bt]$, $[tb]$, and $[bb]$ regions.
Fig.~\ref{fig:reject} summarizes the results of this uncertainty-based rejection strategy.
In panel (a), we show a calibration plot, where accuracy is evaluated across different bins of predictive uncertainty $H(\bar{p})$.
All three curves approximately follow the diagonal, indicating that the model’s uncertainty estimates are well-calibrated and meaningfully reflect the true likelihood of correctness.

Panel (b) displays the accuracy-rejection curve, where predictions are progressively filtered by removing the most uncertain samples.
As the rejection threshold increases, the accuracy of the retained predictions steadily improves, demonstrating that uncertainty estimates can indeed serve as a reliable indicator of prediction quality.

For completeness, we replicate this analysis using the model’s confidence score, defined as $\max_c \bar{p}_c$, in Appx.~\ref{appx:confidence}, and observe a qualitatively similar trend.

One has to note that the model’s predictive uncertainty is not formally equivalent to the $\sigma$-intervals commonly used in high-energy physics to define confidence levels, but importantly, our results demonstrate that it is nonetheless well-calibrated in an operational sense. 
Our results showcase that uncertainty estimates derived from posterior sampling can be effectively used as post-hoc filters, allowing practitioners to systematically trade prediction coverage for reliability.

\section{Inference on Experimental Data}
\label{sec:results}
The primary application of the developed framework is to infer the pole structure of experimentally observed line shapes.
We now demonstrate its use on a dataset of $5 \times 10^3$ measured line shapes, processed using the feature extraction pipeline introduced in Sec.~\ref{sec:features}.
We apply the $C3re$ model, trained on synthetic data as described in Sec.~\ref{sec:model-selection}, to infer the pole configurations for the experimental samples.
The relationship between predicted class distributions, their associated probabilities, and uncertainties is visualized in Fig.~\ref{fig:exp}.

For each of the three predicted pole types ($[bt]$, $[bb]$, and $[tb]$) we sort the predictions by increasing uncertainty (x-axis) and display the corresponding class probability vectors (y-axis).
Each subplot shows stacked, filled areas indicating the predicted class probabilities per sample, overlaid with a dashed line representing the model’s predictive uncertainty.
Confident predictions (left side of each panel) exhibit more peaked distributions largely dominated by a single class, while uncertain predictions (right side) show more diffuse or multi-modal distributions.
The model predicts the configurations $[bt]=[1]$, $[bb]=[2]$, and $[tb]=[1]$ with high confidence.
At the highest confidence, the model assigns class probabilities of $0.97$, $0.93$, and $0.98$, respectively.
The predictive uncertainties for these dominant predictions are also low, further supporting their reliability.
Based on the analysis in Sec.~\ref{sec:reject}, we interpret these results with high model-based confidence, suggesting that the inferred pole structure is consistent with the model’s internal uncertainty calibration.
Alternatively, the model's predictive uncertainty can be used to support interpretation from a different perspective: 
predictions assigning two poles to the $[bt]$ sheet occur predominantly in high-uncertainty regions, and can therefore be deemed unreliable and rejected.

This result differs from the analysis in Ref.~\cite{santosPoleStructure$P_psi^N4312^+$2024}, which favored a configuration of $[bt]=[1]$, $[bb]=[1]$, and $[tb]=[1]$ for the $P_{c\bar{c}}(4312)^+$ state. 
Crucially, the referenced study restricts the total pole count to $n \leq 3$, due to its use of a fully discretized label space with polynomial scaling in $n$. 
In contrast, our approach combines improved feature extraction with a more flexible model formulation, allowing for predictions that include higher-order pole configurations, in our case, up to $n=4$, and thus captures structural variations that were previously inaccessible.

The predicted pole structure can now serve as a guide in the formulation of a dynamical model. 
The real part of the $[bt]$ pole lies below the $\Sigma_c^+\bar{D}^0$ threshold, as it is the only extracted pole capable of directly producing the enhancement observed in Fig.~\ref{fig:lhcb}. 
Moreover, the absence of a peaking structure above the $\Sigma_c^+\bar{D}^0$ threshold implies that the two $[bb]$ poles must also reside below this threshold. 
The $[bt]$ pole together with one of the $[bb]$ poles resemble a pole-shadow pair, which can be associated with a genuine resonance decaying into the $J/\psi p$ channel. 
In contrast, the remaining $[tb]$ pole and the other $[bb]$ pole form a shadow-pole pair characteristic of a virtual state pole with width~\cite{Ikeda:2011dx} associated with the higher $\Sigma_c^+\bar{D}^0$ channel.

\begin{figure}[t!]
    \centering
 \includegraphics[width=0.85\linewidth]{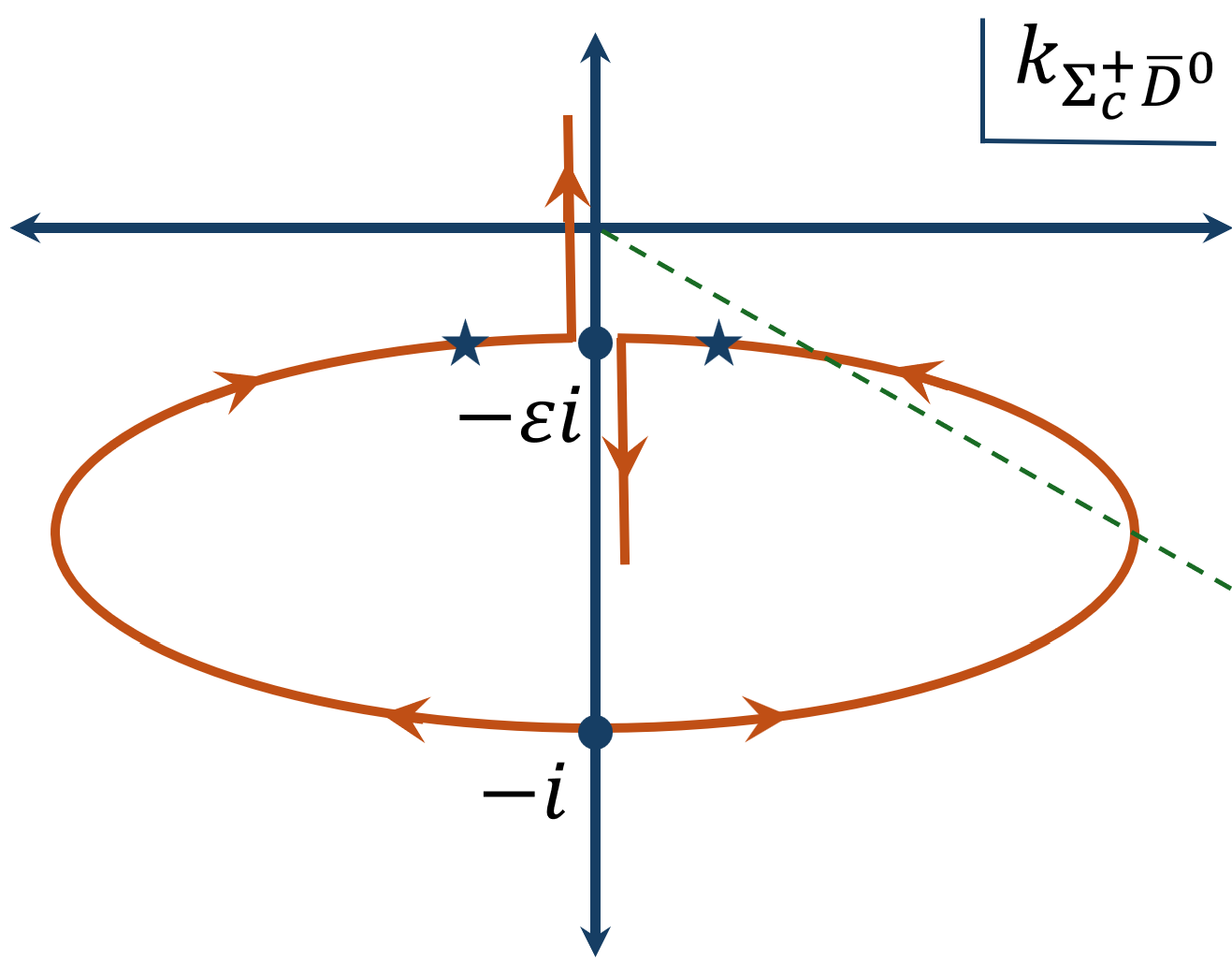}
    \caption{Pole trajectory in the single $\Sigma_c^+\bar{D}^0$ channel. The dashed line corresponds to the line $\Re{k}_{\Sigma_c^+\bar{D}^0}+\Im{\Sigma_c^+\bar{D}^0}=0$. The virtual state with width pole ($\star$) will produce a pole-shadow pair in the $[bb]$ and $[tb]$ sheets when coupled to the lower mass $J/\psi p$ channel.
    } 
    \label{fig:eps_pos}
\end{figure}

In the zero channel coupling limit, such a virtual state pole with finite width can arise from an energy-dependent single-channel interaction, analogous to the Weinberg–Tomozawa contact term~\cite{Ikeda:2011dx,Bruns:2010sv}. 
For instance, in a separable-type interaction, the potential may take the form 
\begin{equation}
    v(p, p')=\lambda(E-M)
   \dfrac{\Lambda^2}{p^2+\Lambda^2}\dfrac{\Lambda^2}{p'^2+\Lambda^2}
    \label{eq:separable}
\end{equation}
where $E = p^2/2\mu + m_1 + m_2$ is the on-shell scattering energy, $\lambda$ is the coupling constant, $\Lambda$ is a cut-off parameter, and $M$ is a mass parameter. The resulting scattering amplitude will produce nearby poles which can manifest as enhancements in the scattering data. We use the following dimensionless quantities to make the discussion more general: $\eta=\lambda\pi\Lambda^3/4$, $\varepsilon=2\mu(\epsilon_{\Sigma_c^+\bar{D}^0}-M)/\Lambda^2$, and $k_{\Sigma_c^+\bar{D}^0} = p/\Lambda$.
As $\eta$ is tuned more negative from zero, the scattering amplitude develops a pole that traces the trajectory shown in Fig.~\ref{fig:eps_pos}. 
If the coupling becomes sufficiently strong, this pole can cross the threshold $k_{\Sigma_c^+\bar{D}^0} = 0$, producing a bound state. 
However, the pole structure extracted in our present analysis indicates that the $\Sigma_c^+\bar{D}^0$ interaction strength is insufficient to form a bound state, and instead gives rise to a virtual state pole with width. 
When coupled to the $J/\psi p$ channel, this manifests as a pole-shadow pair located in the $[bb]$ and $[tb]$ sheets.
Finally, we leave models trained with pole configurations involving total counts $n > 4$ to future studies, as the current analysis yields predictions with sufficiently low total uncertainty for the specific experimental data considered.

\section{Conclusion and Outlook}
\label{sec:conclusion}
A central goal in hadron spectroscopy is to identify and characterize exotic hadronic states emerging near two-particle thresholds. 
These near-threshold enhancements often admit multiple competing interpretations, ranging from hadronic molecules and compact multiquark states to purely kinematical effects. 
Disentangling these scenarios requires new approaches that go beyond traditional amplitude analysis and can rigorously handle the inherent ambiguity of the data.

In this work, we introduced a machine learning framework for inferring pole structures in coupled-channel scattering amplitudes, with a particular focus on quantifying and leveraging predictive uncertainty. 
By combining problem-inspired feature extraction with ensemble-based classifier chains, our approach accurately identifies the underlying configuration of $S$-matrix poles across different Riemann sheets. 
We demonstrated that incorporating predictive uncertainty not only improves classification reliability through confidence-based sample rejection, but also provides deeper insight into the ambiguity inherent in near-threshold phenomena.

Using synthetic data generated from a model-agnostic, analytically tractable $S$-matrix, we demonstrated that the framework can resolve subtle differences between pole configurations across Riemann sheets. 
Importantly, we further applied our trained model to experimental data associated with the $P_{c\bar{c}}(4312)^+$ enhancement and identified a pole configuration of $[bt]=[1],[bb]=[2],[tb]=[1]$ with low uncertainty, consistent with the presence of a compact hidden-charm resonance.
The framework’s uncertainty estimates allowed us to reject alternative interpretations above $95\%$ confidence.
This demonstrates the utility of our method for model-independent interpretation of exotic hadronic signals, where conventional fitting procedures may fail due to degeneracies in line shapes.

Looking ahead, this framework opens several promising directions. 
Extensions to more complex multi-channel systems and higher-dimensional label spaces could widen the method’s applicability to a broader class of near-threshold phenomena. 
Integrating complementary experimental observables, such as angular distributions or other decay channels, may further reduce ambiguities and strengthen physical interpretations. 
Lastly, incorporating active learning strategies could guide data acquisition toward regions of high epistemic uncertainty, optimizing both model training and interpretation.

Overall, this work lays the foundation for uncertainty-aware, data-driven spectroscopy of hadronic resonances, offering a scalable and interpretable alternative to traditional amplitude analysis.

\begin{acknowledgments}
F.F., E.vN, and P.E. thank members of aQa Leiden for fruitful discussions.
F.F., E.vN, and P.E. acknowledge the support received from the Dutch National Growth Fund
(NGF), as part of the Quantum Delta NL programme. 
P.E. acknowledges the support received through the NWO-Quantum Technology program (Grant No.~NGF.1623.23.006) and funding by the Carl-Zeiss-Stiftung (CZS Center QPhoton). 

This work was initiated during the long-term workshop HHIQCD2024 at the Yukawa Institute for Theoretical Physics, Kyoto University (YITP-T-24-02). P.E. and D.L.B.S. gratefully acknowledge the workshop organizers and participants for a stimulating and productive environment.  

Views and opinions expressed are those of the author(s) only and do not necessarily reflect those of the funding institutions. 
Neither of the funding institutions can be held responsible for them.

Parts of this work were performed using the compute
resources from the Academic Leiden Interdisciplinary Cluster Environment (ALICE) provided by Leiden University.

The code used for the computations and the data obtained will be made available online upon publication.
\end{acknowledgments}
\bibliography{felixBib}

\appendix

\section{Training Details}
\label{appx:training}
In this manuscript, we implement the Gradient Boosting classification models using the CatBoost Python library~\cite{dorogushCatBoostGradientBoosting2018} 
and all models are trained using the \texttt{CatBoostClassifier}.
The loss function is set to \texttt{MultiClass}, which is appropriate for our multiclass classification setting and optimizes the cross-entropy loss across all classes. 
We train for up to 1000 boosting iterations, a value that offers a good trade-off between model capacity and computational cost. 
The learning rate is fixed at 0.03, which empirically resulted in stable convergence during model training. 
Each tree in the ensemble has a maximum depth of 6, a commonly used setting that balances model expressiveness and generalization. 
We employ early stopping with a patience of 15 rounds on a held-out validation set, allowing training to terminate automatically if no improvement in validation loss is observed. 
This prevents unnecessary computation and helps mitigate overfitting.

\section{Motivating Cumulative Uncertainty}
\label{appx:cascading}

In Sec.~\ref{sec:uncertainty}, the use of the $C3re$ model was motivated by two key advantages of the classifier chain architecture:
First, $C3$-style models enable per-pole-type predictions, allowing for meaningful uncertainty estimates.
Second, the chain structure allows for correlated prediction across pole-types, leading to improved overall performance (see Fig.~\ref{fig:clf-performance}).

However, a natural drawback of this design is that errors can also become correlated.
Specifically, a misprediction in an early stage of the chain may propagate forward and negatively influence subsequent predictions; a phenomenon we refer to as cascading errors.
This motivates the use of cumulative uncertainty as a diagnostic tool to better track and interpret prediction reliability across the entire chain.

We illustrate this issue using the predictions shown in Fig.~\ref{fig:uncer}.
In particular, the accuracy for class 4 in the $[tb]$ sheet is notably low.
Upon closer inspection, this anomaly is attributable to cascading errors: a failure in earlier stages of the classifier chain leads to compounding mistakes in the final prediction.

This error propagation is detailed in Table~\ref{tab:cascade_example}. 
The true label for the example indicates no poles in $[bt]$ and $[bb]$, and four poles in $[tb]$.
However, the model incorrectly predicts spurious poles in $[bt]$ and $[bb]$, and these incorrect outputs are then passed to the third model, which ultimately predicts class 0 for $[tb]$; a confident but incorrect outcome.

Crucially, although the individual uncertainty for $[tb]$ may appear low, the cumulative uncertainty across all predictions is still high.
Thus, observing the cumulative uncertainty measures $\bar{ \mathcal{H}}_{[bt]} \coloneqq \mathcal{H}_{[bt]}$, $\bar{ \mathcal{H}}_{[bb]}\coloneqq\mathcal{H}_{[bt]}+\mathcal{H}_{[bb]}$, and $\bar{ \mathcal{H}}_{[tb]}\coloneqq\mathcal{H}_{[bt]}+\mathcal{H}_{[bb]}+\mathcal{H}_{[tb]}$ successfully flags the prediction as unreliable, capturing the compounded effect of upstream errors.
In contrast, relying solely on individual uncertainties can obscure such failure modes.
This underscores the importance of cumulative uncertainty in reliably identifying problematic predictions in classifier chain models.

\begin{table}[h]
\centering
\renewcommand{\arraystretch}{1.2}
\begin{tabular}{@{}l@{}}
\toprule
\textbf{Cascaded Prediction Breakdown for [bt], [bb], [tb]} \\
\midrule
\texttt{True labels}:\quad [0], [0], [4] \\
\texttt{Predicted}:\quad [3], [1], [0] \\
\texttt{Entropies}:\quad $[0.68 \pm 0.15],\;[ 0.09 \pm 0.17],\; [0.03 \pm 0.07]$ \\
\texttt{Confidences}:\quad $[0.59 \pm 0.12],\; [0.97 \pm 0.09],\; [0.99 \pm 0.02]$ \\
Error in [bt] propagates → mispredictions in [bb] and [tb] \\
\bottomrule
\end{tabular}
\caption{Example of cascading errors in classifier chains: mispredictions in $[bt]$ and $[bb]$ propagate to the final stage, leading to an overconfident but incorrect prediction for $[tb]$. This motivates introducting a notion of cumulative uncertainty to identify such failure cases.}
\label{tab:cascade_example}
\end{table}

\section{Regression Approach}
\label{appx:regression}
In Sec.~\ref{sec:class_vs_reg}, we introduced the learning task of predicting the non-negative integer pole counts $k_i \in \mathbb{Z}_{\geq 0}$ subject to the constraint
\begin{equation}
    \sum_{i=1}^{s} k_i \leq n,
\end{equation}
where $s$ denotes the number of distinct Riemann sheets (or pole types), and $n$ specifies the maximum total pole count. 
In the main text of this manuscript, we framed this task as a classification problem by one-hot encoding either the full pole configuration or the per-position pole counts.
A natural alternative representation of this framework is to treat the problem as a multivariate regression task, where the labels are given by vectors $\mathbf{y} \in \mathbb{Z}^s$, and the model directly predicts the pole counts at each sheet. 
In the case considered in this manuscript, with $s = 3$, the label might be, for example, $\mathbf{y} = (1, 1, 0)$. 
The key advantage of this formulation is that the label dimensionality scales linearly with $s$, and is independent of the maximum pole count $n$.
However, this approach presents a different challenge: 
The choice of the loss function when training a machine learning model implicitly assumes properties of the underlying distribution of the learning targets~\cite{princeUnderstandingDeepLearning2023}.
The standard loss for regression is the mean squared error, which assumes Gaussian-distributed labels and is designed for continuous output spaces. 
This may not be appropriate for our learning goal of predicting discrete integers $k_i \in \mathbb{Z}_{\geq 0}$.

To assess the feasibility of this approach, we train a multivariate Gradient Boosting regression model to predict labels in $\mathbb{Z}^s$. 
Integer-valued predictions are obtained by rounding the model's continuous outputs. 
The resulting 5-fold validation accuracy of $63\%\pm2\%$ is lower than that of the classification-based models shown in Fig.~\ref{fig:clf-performance}. 
This suggests that, despite the attractive scaling properties, the regression formulation suffers from a misalignment between the label structure and the assumptions of the regression loss.
We leave possible further refinements to adapt the learning task to regression-style models for future work.

\section{Confidence Based Rejections}
\label{appx:confidence}
\begin{figure}[t!]
    \centering
 \includegraphics[width=\linewidth]{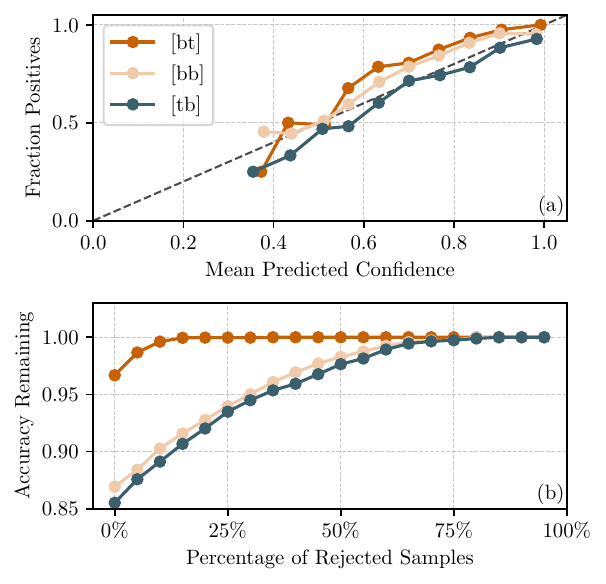}
    \caption[]{\textbf{Prediction-Rejection Ratio.}
The $C3re$ ensemble model is trained to predict the pole structures $[bt]$, $[tb]$, and $[bb]$.
The figure shows two complementary visualizations for evaluating the model confidence.
(a) Calibration plot showing the fraction of correct predictions as a function of the predicted confidence $\max_c \bar{p}_c$.
All three curves follow the linear trend on the diagonal, meaning that the the model confidences reflect actual accuracy, indicating well-calibrated classifier.
(b) Accuracy-rejection curve based on cumulative predicted confidence.
Accuracy is computed after discarding an increasing percentage of low-confidence samples.
As more low confidence predictions are rejected, accuracy of the retained predictions steadily improves, confirming that the model’s confidence estimates are linked to correctness and can serve as a reliable rejection criterion.
}\label{fig:appx-reject}
\end{figure}

In Sec.~\ref{sec:reject}, we explored the use of predictive uncertainty as a rejection criterion to systematically filter out unreliable predictions.
In this, uncertainty was quantified via the entropy of the predicted class probabilities, as defined in Eq.~\eqref{eq:uncertainty}.

Here, we extend this analysis by evaluating an alternative and widely used criterion: model confidence, defined as the maximum predicted class probability, $\max_c \bar{p}_c$.
While entropy captures the overall uncertainty across all classes, confidence focuses solely on the most likely prediction.
We therefore assess whether this metric can also serve as an effective basis for selective prediction.

Following the same procedure as in Sec.~\ref{sec:reject}, we define a range of confidence thresholds and evaluate model performance only on those samples whose confidence exceeds the given threshold.
If model confidence is well-calibrated, then higher-confidence predictions should, on average, be more accurate.
Conversely, discarding low-confidence samples should improve the reliability of the retained predictions~\cite{lakshminarayananSimpleScalablePredictive2017}.

Again, the $C3re$ ensemble model is trained to classify the presence of poles in the $[bt]$, $[tb]$, and $[bb]$ regions.
Fig.~\ref{fig:appx-reject} illustrates the confidence-based rejection strategies:
Panel (a) shows a calibration plot using model confidence $\max_c \bar{p}_c$, where accuracy is evaluated across different confidence bins.
All three curves approximately follow the diagonal, indicating that the model confidences are well-calibrated and informative.
Panel (b) presents the accuracy-rejection curve using confidence as the rejection criterion.
As low-confidence predictions are progressively removed, the accuracy of the retained predictions improves steadily, mirroring the trend observed with uncertainty-based rejection in Fig.~\ref{fig:reject}.

These results confirm that both uncertainty and confidence can be used as post-hoc filtering mechanisms, enabling practitioners to trade off prediction coverage for reliability.
However, while confidence is easier to compute and interpret, it lacks the richer distributional information provided by entropy. 
The choice between them should therefore depend on the requirements of the downstream application.

\end{document}